\documentclass[twocolumn,showpacs,preprintnumbers,amsmath,amssymb]{revtex4-1}

\usepackage{graphicx}
\usepackage{dcolumn}
\usepackage{bm}
\usepackage{booktabs}
\usepackage{amsmath,amssymb,mathrsfs}

\begin{document}


\def\preprint{0}

\title{Free electron lasers assisted by pseudo energy bands}

\author{Takashi Tanaka}\email{ztanaka@spring8.or.jp}
\author{Ryota Kinjo}

\affiliation{%
RIKEN SPring-8 Center, Koto 1-1-1, Sayo, Hyogo 679-5148, Japan
}%

\date{\today}

\begin{abstract}
We propose a new scheme for high gain harmonic generation free electron lasers (HGHG FELs), which is seeded by a pair of intersecting laser beams to interact with an electron beam in a modulator undulator located in a dispersive section. The interference of the laser beams gives rise to a two-dimensional modulation in the energy-time phase space because of a strong correlation between the electron energy and the position in the direction of dispersion. This eventually forms pseudo energy bands in the electron beam, which results in efficient harmonic generation in HGHG FELs in a similar manner to the well-known scheme using the echo effects, with the requirement on the energy modulation being much more relaxed.
\end{abstract}

\maketitle

Free electron lasers (FELs) have now become a promising scheme to produce high-power coherent light in short wavelength regions which are not accessible with conventional lasers. In spectral regions where reflective mirrors are not available, FELs currently under operation are divided into two types based on different principles: self amplified spontaneous emission (SASE) \cite{Kondratenko-PA-1980} and high gain harmonic generation (HGHG) \cite{Yu-PRA-1991}. The former amplifies spontaneous radiation, while the latter upconverts coherent seed light, or seed laser. Although the latter has many advantages against the former, its attainable wavelength is technically limited by two factors: wavelength ($\lambda_s$) of the available seed laser, and upconversion efficiency of the HGHG process. 

Shortening $\lambda_s$ has been explored for many years; even so, that of the seed laser currently available in HGHG FELs is in the spectral region from infrared to ultraviolet, mainly because it should be sufficiently stable to be synchronized with the electron beam spatiotemporally. It is thus the recent trend to improve the upconversion efficiency to realize HGHG FELs in shorter wavelengths.

The upconversion in HGHG FELs is achieved by upconverting the microbunch formed in the electron beam through interaction with the seed laser, which makes it possible to generate coherent light having the wavelength of $\lambda_s/n$, where $n$ is an integer referred to as a harmonic order. It is well known that the electron beam should be strongly modulated in energy with an amplitude $\gamma_M$ larger than or comparable to $n\sigma_{\gamma}$, where $\sigma_{\gamma}$ is the RMS energy spread of the electron beam; this requirement poses two issues in HGHG FELs to aim at higher $n$. First, the required power of the seed laser, which is roughly proportional to $\gamma_M^2$, increases as $n^2$. Second, which may be more critical, larger $\gamma_M$ eventually deteriorates the electron beam quality and thus limits the achievable FEL performances.

To overcome the above difficulty in HGHG FELs, two different schemes based on special schemes for energy modulation have been proposed, which are referred to as the echo-enabled harmonic generation (EEHG) \cite{Stupakov-PRL-2009} and the phase-merging enhanced harmonic generation (PEHG) \cite{Deng-PRL-2013}. The electron beam is modulated twice in the former, while the modulation is applied in a dispersive section in the latter; both of them significantly relax the above requirement on the energy modulation and thus the upconversion efficiency is much better than the original HGHG scheme. An attractive application of these schemes is to generate coherent radiation in storage rings (SRs) especially in short-wavelength regions \cite{Evain-NJP-2012,Feng-OSAProc-2016}, which significantly enhances the achievable brightness of existing and upcoming SR facilities. As discussed later in detail, however, the quality of the electron beam can be significantly deteriorated in these schemes once it is modulated by the seed laser. This effectively limits the achievable performances in SRs where the electron beam should be repeatedly used. In this Letter, we propose a new scheme for HGHG FELs, which can retard the quality deterioration of the electron beam, and thus is well compatible with the SR operation.

Figure 1 shows a schematic layout of the proposed scheme, with the coordinate system to be used in the following discussions. It is composed of three undulators similar to the EEHG scheme, which are referred to as the first modulator, second modulator, and radiator. The difference is that the first modulator is located in a dispersive section, and a pair of lasers intersecting at an angle of $2\alpha$ work as the seed laser.

\begin{figure}[htb]
\centering
\includegraphics[width=\linewidth]{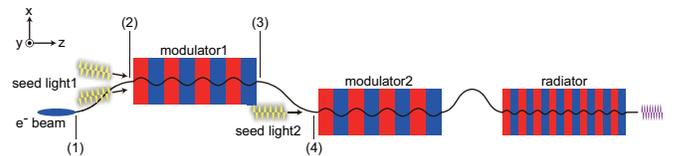}
\caption{Layout of the proposed scheme; the numbers from (1) to (4) show the locations where the electron distributions are illustrated in Fig. 2.}
\label{fig:fig1}
\end{figure}

Let us first consider the energy exchange between the electron beam and a pair of intersecting laser beams (ILBs) in the 1st modulator. The electric field of the ILBs is given as
\[
E(x,z,s)=2E_0\cos(\alpha k_sx+\phi/2)\mbox{e}^{i(k_ss-k_uz)},
\]
with $k_s=2\pi/\lambda_s$ and $k_u=k(c/\overline{v}_z-\cos\alpha)$, where $s$ is the relative longitudinal position with respect to the electron beam center, $\overline{v}_z$ is the average longitudinal velocity of the electron beam moving in the modulator, $c$ is the speed of light, $E_0$ is the field amplitude of the seed laser, and $\phi$ is the relative phase between the ILBs and assumed to be $\pi$ in the following discussion. The energy modulation $\delta\gamma$ given in the electron beam is then given by
\[
\delta\gamma(x,s)=\gamma_M\sin(\alpha k_s x)\cos(k_ss),
\]
where $\gamma_M$ depends linearly on $E_0$ and the 2nd factor means that the energy modulation is a function of $x$ as well as $s$, which comes from the interference of the ILBs. To facilitate the following discussion, we introduce a variable $\eta=\gamma/\gamma_0-1$, where $\gamma$ is the Lorentz factor of an electron and $\gamma_0$ denotes its average over the electron beam.

We now discuss the motion of electrons in the $(x,\eta)$ phase space while they travel through the dispersive section with the strength $D$, where the 1st modulator is located. To be specific, we consider the electron distributions at four different locations $(1)\sim(4)$ indicated in Fig. 1. We assume an electron initially positioned at $x=0$ and $\eta=\eta'$ before entering the dispersive section, and define $x_{m}$ and $\eta_{m}$ as the coordinates at the $m$-th position. Then it is easy to show $x_{1}=0,\:\: x_{2}=x_{3}=\eta'D,\:\: x_{4}=-\eta_MD\sin(\pi\eta'/2\eta_{M0})\cos(k_ss)$, and $\eta_{1}=\eta_{2}=\eta',\:\: \eta_{3}=\eta_{4}=\eta'+\eta_M\sin(\pi\eta'/2\eta_{M0})\cos(k_ss)$, with $\eta_M=\gamma_M/\gamma_0$ and $\eta_{M0}=\pi/2\alpha k_sD$. 

The evolution of the electron distribution mathematically given above is illustrated in Fig. 2 in two particular cases of $s=0$ and $s=\lambda_s/2$, where the numbers correspond to respective longitudinal positions. Note that no difference is found between the two cases in (1) and (2) before the electron beam enters the 1st modulator. As found in Fig. 2-(4), the electrons after being modulated by the ILBs are more populated around the energies indicated by arrows, if $\eta_M$ is optimized. In other words, the energy distribution of the electron beam is quantized at discrete levels separated by an energy interval of $\Delta\eta=2\eta_{M0}$.

\begin{figure}[htb]
\centering
\includegraphics[width=\linewidth]{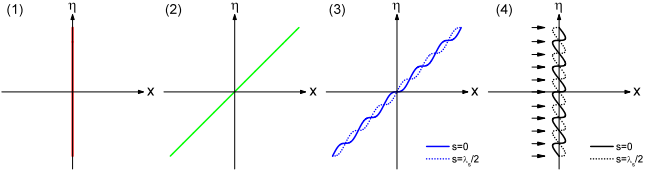}
\caption{Electron distributions in the phase space at four different locations indicated in Fig. 2. The solid and dashed lines in (3) and (4) correspond to $s=0$ and $s=\lambda_s/2$, respectively.}
\label{fig:fig2}
\end{figure}

It is obvious that the above effects can be more or less smeared under realistic conditions, especially by the beam size in the direction of dispersion, and the energy levels may transfer to energy bands. In practice, such a structure should be referred to as ``pseudo energy bands'', because it has nothing to do with quantum mechanics. If the beam size is too large, the band structure can completely disappear and the original distribution function is restored. To quantify its impact, let us evaluate the energy distribution function $F(\eta)$ at the position (4), assuming that the energy and horizontal distribution functions of the electron beam are given by Gaussian functions with the standard deviations of $\sigma_{\eta}$ and $\sigma_x$. In the same manner as described above, the coordinate of an electron initially positioned at $(x',\eta')$ is converted as $x_4=x'+(\eta'-\eta)D$ and $\eta_4=\eta'+\eta_M\sin[(\pi/2\eta_{M0})(x'/D+\eta')]\cos(k_ss)$. Then we have

\ifnum \preprint=0
\begin{eqnarray*}
\hat{F}(\hat{\eta})&=&\frac{1}{2\pi^2\hat{\sigma}_x}\int_{-\pi/2}^{\pi/2}d\psi\int_{-\infty}^{\infty}d\hat{x} \\
&\times&\exp\left\{-\frac{[\hat{x}-H(\hat{\eta},\hat{x},\psi)]^2}{2\hat{\sigma}_x^2}-\frac{H^2(\hat{\eta},\hat{x},\psi)}{2}\right\},
\end{eqnarray*}
\else
\[
\hat{F}(\hat{\eta})=\frac{1}{2\pi^2\hat{\sigma}_x}\int_{-\pi/2}^{\pi/2}d\psi\int_{-\infty}^{\infty}d\hat{x}\exp\left\{-\frac{[\hat{x}-H(\hat{\eta},\hat{x},\psi)]^2}{2\hat{\sigma}_x^2}-\frac{H^2(\hat{\eta},\hat{x},\psi)}{2}\right\},
\]
\fi

with
\[
H(\hat{\eta},\hat{x},\psi)=\hat{\eta}+\hat{\eta}_M\sin\psi\sin(\pi\hat{x}/2\hat{\eta}_{M0}),
\]
where we have introduced normalized variables defined by $\hat{x}=x/\sigma_{\eta}D$, $\hat{\eta}=\eta/\sigma_{\eta}$, $\hat{\sigma}_x=\sigma_x/\sigma_{\eta}D$, $\hat{\eta}_M=\eta_M/\sigma_{\eta}$, $\hat{\eta}_{M0}=\eta_{M0}/\sigma_{\eta}$, and $\hat{F}=\sigma_{\eta}F$.

Figure 3(a) shows the plots of $\hat{F}(\hat{\eta})$ computed with $\hat{\eta}_{M}=\hat{\eta}_{M0}=0.1$, for different values of $\hat{\sigma}_x$, with inset showing the detail around $\hat{\eta}=0$. A number of peaks, or more specifically energy bands, are found as expected, whose intensities are sensitive to $\hat{\sigma}_x$. For example, $\hat{F}(\hat{\eta})$ reduces to the original Gaussian function when $\hat{\sigma}_x=0.1$. It is thus important to have a smaller beam size, or a smaller emittance, in the direction of dispersion to generate a more distinct energy-band structure.

\begin{figure}[htb]
\centering
\includegraphics[width=\linewidth]{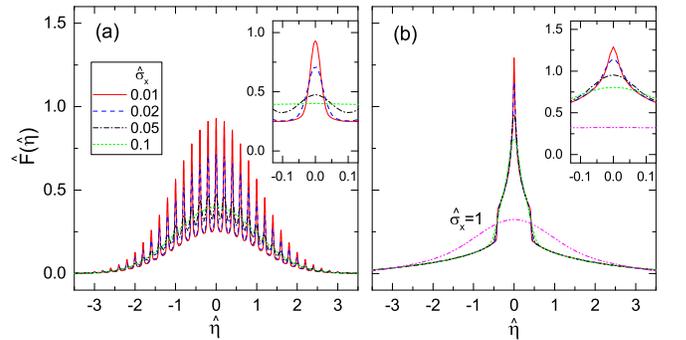}
\caption{Energy distribution functions $\hat{F}(\hat{\eta})$ for different conditions of $\hat{\sigma}_x$: (a) $\hat{\eta}_{M0}=0.1$ and (b) $\hat{\eta}_{M0}=2$. In both cases, $\hat{\eta}_{M}=\hat{\eta}_{M0}$.}
\label{fig:fig3}
\end{figure}

We now note that 10 $(=\hat{\eta}_{M0}^{-1})$ energy bands are found within the region of $|\hat{\eta}|\leq 1$; this is obvious from the fact that the energy-band interval is given as $2\hat{\eta}_{M0}$. In this regard, it is interesting to decrease the number of energy bands by applying larger $\hat{\eta}_{M0}$. For example, $\hat{F}(\hat{\eta})$ for different values of $\hat{\sigma}_x$ with $\hat{\eta}_{M0}=2$ are plotted in Fig. 3(b), where the only energy band appears around $\hat{\eta}=0$ with a higher peak intensity, meaning that the energy spread of the electron beam is effectively reduced. This suggests another possibility of the proposed scheme, which is to be discussed elsewhere.

After the energy bands are formed in the electron beam in the 1st modulator, it further undergoes an energy modulation in the 2nd modulator, which is then converted to a density modulation in the following chicane with the momentum compaction of $R_{56}$. To discuss the modulation there, let us assume that the distribution function of the electron beam in the $(s,\eta)$ phase space, before being injected to the 2nd modulator, is approximately given by
\[
G(s,\eta)=\frac{1}{\sqrt{2\pi}\sigma_{\eta}}\exp\left(-\frac{\eta^2}{2\sigma_{\eta}^2}\right)\left[1+\delta_{F}\cos\left(\frac{\pi\eta}{\sigma_{\eta}\hat{\eta}_{M0}}\right)\right],
\]
where $\delta_F$ denotes the population of electrons contained in the formed energy bands. The validity of this approximation is easily verified by comparing with Fig. 3(a). Note that $G(s,\eta)$ does not actually include the variable $s$, meaning that no density modulation is present after passing through the dispersion section because of the large momentum compaction. After the electron beam passes through the 2nd modulator and the chicane, the distribution function is modified according to the energy and density modulations, and is given by replacing $\eta$ in $G(s,\eta)$ by $\eta-\eta_H\sin[k_s(s-R_{56}\eta)]$, where $\eta_H$ is the energy modulation amplitude given in the 2nd modulator. The bunching factor $b_n$ at the $n$-th harmonic is then given as
\[
b_n\equiv \frac{1}{\lambda_s}\int_{-\lambda_s/2}^{\lambda_s/2}ds\int_{-\infty}^{\infty}d\eta\:\:G(s,\eta)=b_{1n}+b_{2n},
\]
with
\[
b_{1n}=J_n(n\rho\hat{\eta}_H)\exp(-n^2\rho^2/2),
\]
and
\ifnum \preprint=0
\begin{eqnarray*}
b_{2n}&=&\frac{\delta_FJ_n(n\rho\hat{\eta}_H)}{2}\left\{
\exp\left[-\frac{1}{2}\left(n\rho+\frac{\pi}{\hat{\eta}_{M0}}\right)^2\right] \right. \\
&+& \left. \exp\left[-\frac{1}{2}\left(n\rho-\frac{\pi}{\hat{\eta}_{M0}}\right)^2\right]
\right\},
\end{eqnarray*}
\else
\[
b_{2n}=\frac{\delta_FJ_n(n\rho\hat{\eta}_H)}{2}\left\{
\exp\left[-\frac{1}{2}\left(n\rho+\frac{\pi}{\hat{\eta}_{M0}}\right)^2\right]+\exp\left[-\frac{1}{2}\left(n\rho-\frac{\pi}{\hat{\eta}_{M0}}\right)^2\right]
\right\},
\]
\fi
where $\rho=k_sR_{56}\sigma_{\eta}$ is the normalized momentum compaction of the chicane. The 1st term $b_{1n}$ denotes the bunching factor due to the normal HGHG process exponentially decaying as $n^2$, while the 2nd term comes from the energy-band structure in the electron beam, and reduces to
\[
b_{2n}\sim\delta_FJ_n(\pi\hat{\eta}_H/\hat{\eta}_{M0})/2,
\]
if $\rho=\pi/n\hat{\eta}_{M0}$ is satisfied. It is now obvious that $b_{2n}$ decays as $n$ much more slowly than $b_{1n}$, if relevant parameters are optimized. This suggests that a much larger harmonic order is expected in the HGHG FELs based on the electron beam with the energy-band structure.

It should be mentioned here that the energy modulation in the dispersive section, which is one of the essential processes in the proposed scheme, gives rise to an increase in the beam size and thus the emittance in the direction of dispersion as is evident from Fig. 2-(4). To be specific, the emittance grows by a factor of $\hat{\varepsilon}=\sqrt{1+(\sigma_M/\sigma_x)^2}$, with
\[
\sigma_{M}^2=\left\langle\int_{-\infty}^{\infty} x^2\delta(x-x_4)dx\right\rangle=\left(\frac{D\eta_M}{2}\right)^2,
\]
where $\delta$ is the Dirac delta function and $\langle\ldots\rangle$ means an average with respect to $s$ and $\eta'$. Substituting the normalized variables, we have $\hat{\varepsilon}=\sqrt{1+(\hat{\eta_M}/2\hat{\sigma_x})^2}$. For example, $\hat{\varepsilon}$ amounts to 1.4 in the condition of $\hat{\eta}_M=0.1$ and $\hat{\sigma}_x=0.05$ as shown in Fig. 3(a).

To illustrate a possible performance of the proposed scheme, we need to carry out FEL simulations with realistic electron beam parameters. Here, we assume an electron beam in a SR, instead of a linear accelerator, because it is more compatible with the proposed scheme in terms of a relatively large energy spread and small emittance in the vertical direction. Furthermore, realization of SR-based FELs in short-wavelength regions is apparently attractive in terms of enhancing the peak and average brightness of the SR facilities. 

As an example, we assume a SR with the circumference of 200 m, which stores a single-bunch electron beam with the energy of 1 GeV, natural emittance of 10 nm$\cdot$rad, coupling constant of 0.5\%, energy spread of 0.1\%, bunch charge of 13 nC, and FWHM bunch length of 50 ps. The resultant peak and average beam currents are 250 A and 20 mA, respectively. We also assume that the seed laser with $\lambda_s=267$ nm modulates the electron beam to emit coherent radiation at the 20th harmonic (13.4 nm). The 1st modulator with the period length of 40 mm and peak magnetic field of 1.02 T is located in a vertical dispersive section with $D=0.2$ m, where ILBs with $\alpha=3.34$ mrad and the power density of $5.0\times 10^8$ W/cm$^2$ are injected synchronously with the electron beam.  The 2nd modulator has the period length of 60 mm and peak field of 1.45 T, where the electron beam is further modulated by a single laser beam (not ILB) with the power density of $7.9\times 10^9$ W/cm$^2$. Then, the energy modulation is converted to the density modulation in a chicane with $R_{56}=0.056$ mm, and the electron beam emits coherent radiation at 13.4 nm in the radiator with the period length of 24 mm and peak field of 1.14 T. Note that the (1st and 2nd) modulators  and radiator are 2 m long, and the horizontal and vertical betatron functions are 2 m and 1 m at the center of each of them, respectively. It is relevant to mention that $\hat{\eta}_M\sim\hat{\eta}_{M0}\sim 0.1$ is satisfied with the above parameters, and thus energy bands similar to those shown in Fig. 3(a) are expected.

Note that the time-averaged brightness (given in photons/sec/mm$^2$/mrad$^2$/0.1\%b.w.) of spontaneous radiation at the wavelength of 13.4 nm, which is emitted by the electron beam moving in the radiator, is estimated as 1.9$\times 10^{18}$ with the above parameters. Considering the peak current of 250 A, this can be converted to the peak brightness of 2.4$\times 10^{22}$.

The FEL simulations have been carried out using the simulation code {\it SIMPLEX} \cite{Tanaka-JSR-2015}. Note that only a limited region around the bunch center having the peak current of 250 A has been simulated instead of the whole electron bunch, which is enough to evaluate the expected performance of the proposed scheme.

\begin{figure}[htb]
\centering
\includegraphics[width=\linewidth]{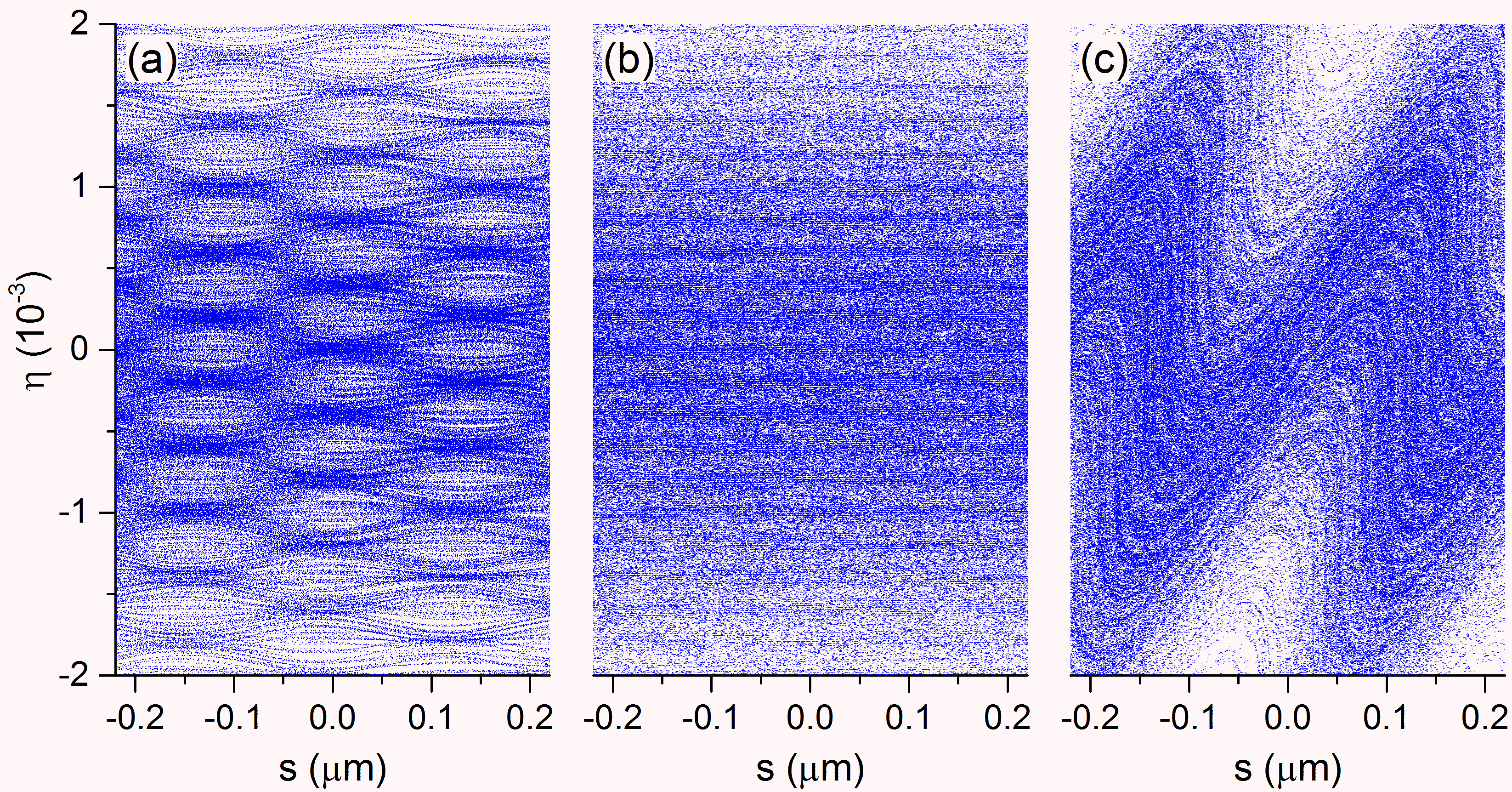}
\caption{Simulation results in terms of the macroparticle distributions: (a) after the 1st modulator, (b) before the 2nd modulator, and (c) before the radiator. }
\label{fig:fig4}
\end{figure}

Figures 4(a)$\sim$(c) show the distributions of macroparticles retrieved from the simulation results after the 1st modulator, before the 2nd modulator, and before the radiator, respectively. The ``web-like pattern'' found in (a) is specific to the modulation by the ILBs, which reduces to the energy bands as found in (b), after the large momentum compaction is applied in the dispersive section. After passing through the 2nd modulator and the chicane, each energy band is independently modulated to generate microbunches with a large harmonic order as found in (c), which is similar to the mechanism of the EEHG scheme. Note that $R_{56}$ of 0.056 mm is an optimum value to maximize the bunch factor at the 20th harmonic; in this example, $|b_{20}|$ is around 0.03. Although this is not a big number, it is large enough to generate intense coherent radiation in the radiator.

Figure 5 shows the growth of coherent radiation with the wavelength of 13.4 nm, where the peak power is plotted as a function of the longitudinal distance from the radiator entrance. At the exit of the radiator, the peak power approaches 100 kW, which corresponds to the peak brightness of 6$\times 10^{29}$ if we assume that the radiation is fully coherent in time and space. This is 7 orders of magnitude higher than that of the spontaneous radiation.

\begin{figure}[htb]
\centering
\ifnum \preprint=0
\includegraphics[width=\linewidth]{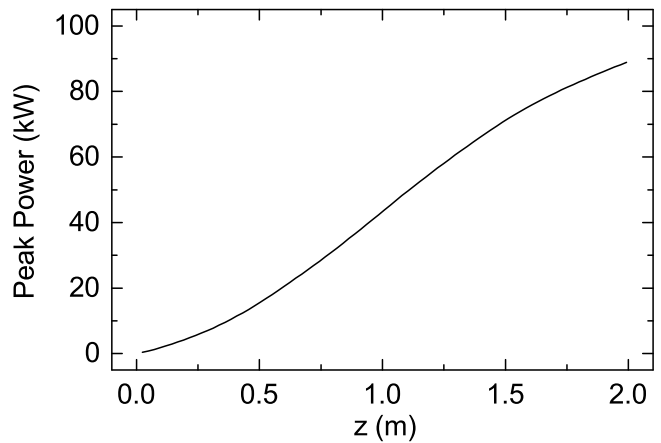}
\else
\includegraphics[width=0.5\linewidth]{fig5.eps}
\fi
\caption{Radiation power growth at the radiator.}
\label{fig:fig5}
\end{figure}

To compute the time-averaged brightness, we should take into account that the electron beam quality is more or less deteriorated in the modulation process. To be specific, the energy spread increases from 0.1\% to 0.11\%, and the vertical emittance from 0.05 to 0.065 nm$\cdot$rad in the above example. We need to wait for the electron beam to recover its equilibrium state through the radiation damping process of the SR. The emittance growth has actually an impact on the achievable performance of the proposed scheme,  and thus we need to wait for $\ln(0.065/0.05)=0.26$ times the damping time, so that the vertical emittance recovers its equilibrium value. Assuming a typical value of a few milliseconds for the damping time, we need to wait for nearly 1 ms after the electron beam is modulated to emit coherent radiation through the proposed scheme. As a result, we can repeat the coherent radiation process using the same electron bunch 1000 times per second. Taking into account the bunch length of 50 ps in this example, the time-averaged brightness reaches $10^{22}$, which is 4 orders of magnitude higher than that of the spontaneous radiation.

Finally, we compare the proposed scheme with other competing schemes, EEHG and PEHG, especially for application to SRs, in which the quality deterioration of the electron beam is a critical issue.

Even though the requirement on the energy modulation amplitude $\eta_M$ is significantly relaxed in the EEHG scheme, it should be still many times larger than the initial energy spread $\sigma_{\eta}$ to achieve a high harmonic order. For example in \cite{Evain-NJP-2012}, which proposes the EEHG-based coherent radiation in a SR, $\eta_M=5\sigma_{\eta}$ is assumed to sufficiently modulate the electron beam, meaning that the energy spread grows by a factor of 3.7.

In the PEHG scheme, the requirement on the energy modulation can be further relaxed, and thus the energy spread growth can be suppressed. It should be noted, however, that the energy modulation in a dispersive section gives rise to the emittance growth as in the case of the proposed scheme. To be specific, $\hat{\varepsilon}$ in the PEHG is roughly given as $\hat{\varepsilon}=\sqrt{1+(\hat{\eta_M}/\sqrt{2}\hat{\sigma_x})^2}$, where the discrepancy in the coefficient ($\sqrt{2}$) comes from the difference in the injection scheme of the seed laser. The energy modulation required in the PEHG scheme can be evaluated by recalling that the finite beam size disturbs the microbunch formation with an effective energy spread of $\sigma_x/D$ \cite{Deng-PRL-2013}, and thus we have $\eta_M\geq n\sigma_x/D$ to reach the harmonic order of $n$, which reduces to $\hat{\varepsilon}\geq\sqrt{1+n^2/2}$. For example, $\hat{\varepsilon}$ amounts to 14.2 for the 20th harmonics.

In comparison to the above two schemes, the quality deterioration of the electron beam in the proposed scheme is expected to be much lower, because the energy modulation to be applied in the dispersive section does not have to be large; in the above example, $\eta_M\sim 0.1\sigma_{\eta}$ is enough to form the pseudo energy bands. In conclusion, the proposed scheme will offer an attractive option in SR-based light sources to significantly enhance the average brightness as well as the peak one.

\bibliography{references}

\begin{thebibliography}{7}%
\makeatletter
\providecommand \@ifxundefined [1]{%
 \@ifx{#1\undefined}
}%
\providecommand \@ifnum [1]{%
 \ifnum #1\expandafter \@firstoftwo
 \else \expandafter \@secondoftwo
 \fi
}%
\providecommand \@ifx [1]{%
 \ifx #1\expandafter \@firstoftwo
 \else \expandafter \@secondoftwo
 \fi
}%
\providecommand \natexlab [1]{#1}%
\providecommand \enquote  [1]{``#1''}%
\providecommand \bibnamefont  [1]{#1}%
\providecommand \bibfnamefont [1]{#1}%
\providecommand \citenamefont [1]{#1}%
\providecommand \href@noop [0]{\@secondoftwo}%
\providecommand \href [0]{\begingroup \@sanitize@url \@href}%
\providecommand \@href[1]{\@@startlink{#1}\@@href}%
\providecommand \@@href[1]{\endgroup#1\@@endlink}%
\providecommand \@sanitize@url [0]{\catcode `\\12\catcode `\$12\catcode
  `\&12\catcode `\#12\catcode `\^12\catcode `\_12\catcode `\%12\relax}%
\providecommand \@@startlink[1]{}%
\providecommand \@@endlink[0]{}%
\providecommand \url  [0]{\begingroup\@sanitize@url \@url }%
\providecommand \@url [1]{\endgroup\@href {#1}{\urlprefix }}%
\providecommand \urlprefix  [0]{URL }%
\providecommand \Eprint [0]{\href }%
\providecommand \doibase [0]{http://dx.doi.org/}%
\providecommand \selectlanguage [0]{\@gobble}%
\providecommand \bibinfo  [0]{\@secondoftwo}%
\providecommand \bibfield  [0]{\@secondoftwo}%
\providecommand \translation [1]{[#1]}%
\providecommand \BibitemOpen [0]{}%
\providecommand \bibitemStop [0]{}%
\providecommand \bibitemNoStop [0]{.\EOS\space}%
\providecommand \EOS [0]{\spacefactor3000\relax}%
\providecommand \BibitemShut  [1]{\csname bibitem#1\endcsname}%
\let\auto@bib@innerbib\@empty
\bibitem [{\citenamefont {Kondratenko}\ and\ \citenamefont
  {Saldin}(1980)}]{Kondratenko-PA-1980}%
  \BibitemOpen
  \bibfield  {author} {\bibinfo {author} {\bibfnamefont {A.}~\bibnamefont
  {Kondratenko}}\ and\ \bibinfo {author} {\bibfnamefont {E.}~\bibnamefont
  {Saldin}},\ }\href@noop {} {\bibfield  {journal} {\bibinfo  {journal}
  {Particle Accelerators}\ }\textbf {\bibinfo {volume} {10}},\ \bibinfo {pages}
  {207} (\bibinfo {year} {1980})}\BibitemShut {NoStop}%
\bibitem [{\citenamefont {Yu}(1991)}]{Yu-PRA-1991}%
  \BibitemOpen
  \bibfield  {author} {\bibinfo {author} {\bibfnamefont {L.~H.}\ \bibnamefont
  {Yu}},\ }\href {\doibase 10.1103/PhysRevA.44.5178} {\bibfield  {journal}
  {\bibinfo  {journal} {Phys. Rev. A}\ }\textbf {\bibinfo {volume} {44}},\
  \bibinfo {pages} {5178} (\bibinfo {year} {1991})}\BibitemShut {NoStop}%
\bibitem [{\citenamefont {Stupakov}(2009)}]{Stupakov-PRL-2009}%
  \BibitemOpen
  \bibfield  {author} {\bibinfo {author} {\bibfnamefont {G.}~\bibnamefont
  {Stupakov}},\ }\href {\doibase 10.1103/PhysRevLett.102.074801} {\bibfield
  {journal} {\bibinfo  {journal} {Phys. Rev. Lett.}\ }\textbf {\bibinfo
  {volume} {102}},\ \bibinfo {pages} {074801} (\bibinfo {year}
  {2009})}\BibitemShut {NoStop}%
\bibitem [{\citenamefont {Deng}\ and\ \citenamefont
  {Feng}(2013)}]{Deng-PRL-2013}%
  \BibitemOpen
  \bibfield  {author} {\bibinfo {author} {\bibfnamefont {H.}~\bibnamefont
  {Deng}}\ and\ \bibinfo {author} {\bibfnamefont {C.}~\bibnamefont {Feng}},\
  }\href {\doibase 10.1103/PhysRevLett.111.084801} {\bibfield  {journal}
  {\bibinfo  {journal} {Phys. Rev. Lett.}\ }\textbf {\bibinfo {volume} {111}},\
  \bibinfo {pages} {084801} (\bibinfo {year} {2013})}\BibitemShut {NoStop}%
\bibitem [{\citenamefont {Evain}\ \emph {et~al.}(2012)\citenamefont {Evain},
  \citenamefont {Loulergue}, \citenamefont {Nadji}, \citenamefont {Filhol},
  \citenamefont {Couprie},\ and\ \citenamefont {Zholents}}]{Evain-NJP-2012}%
  \BibitemOpen
  \bibfield  {author} {\bibinfo {author} {\bibfnamefont {C.}~\bibnamefont
  {Evain}}, \bibinfo {author} {\bibfnamefont {A.}~\bibnamefont {Loulergue}},
  \bibinfo {author} {\bibfnamefont {A.}~\bibnamefont {Nadji}}, \bibinfo
  {author} {\bibfnamefont {J.~M.}\ \bibnamefont {Filhol}}, \bibinfo {author}
  {\bibfnamefont {M.~E.}\ \bibnamefont {Couprie}}, \ and\ \bibinfo {author}
  {\bibfnamefont {A.~A.}\ \bibnamefont {Zholents}},\ }\href
  {http://stacks.iop.org/1367-2630/14/i=2/a=023003} {\bibfield  {journal}
  {\bibinfo  {journal} {New Journal of Physics}\ }\textbf {\bibinfo {volume}
  {14}},\ \bibinfo {pages} {023003} (\bibinfo {year} {2012})}\BibitemShut
  {NoStop}%
\bibitem [{\citenamefont {Feng}\ \emph {et~al.}(2016)\citenamefont {Feng},
  \citenamefont {Jiang}, \citenamefont {Qi},\ and\ \citenamefont
  {Zhao}}]{Feng-OSAProc-2016}%
  \BibitemOpen
  \bibfield  {author} {\bibinfo {author} {\bibfnamefont {C.}~\bibnamefont
  {Feng}}, \bibinfo {author} {\bibfnamefont {B.}~\bibnamefont {Jiang}},
  \bibinfo {author} {\bibfnamefont {Z.}~\bibnamefont {Qi}}, \ and\ \bibinfo
  {author} {\bibfnamefont {Z.}~\bibnamefont {Zhao}},\ }in\ \href {\doibase
  10.1364/EUVXRAY.2016.EM3A.1} {\emph {\bibinfo {booktitle} {High-Brightness
  Sources and Light-Driven Interactions}}}\ (\bibinfo  {publisher} {Optical
  Society of America},\ \bibinfo {year} {2016})\ p.\ \bibinfo {pages}
  {EM3A.1}\BibitemShut {NoStop}%
\bibitem [{\citenamefont {Tanaka}(2015)}]{Tanaka-JSR-2015}%
  \BibitemOpen
  \bibfield  {author} {\bibinfo {author} {\bibfnamefont {T.}~\bibnamefont
  {Tanaka}},\ }\href {\doibase 10.1107/S1600577515012850} {\bibfield  {journal}
  {\bibinfo  {journal} {Journal of Synchrotron Radiation}\ }\textbf {\bibinfo
  {volume} {22}},\ \bibinfo {pages} {1319} (\bibinfo {year}
  {2015})}\BibitemShut {NoStop}%
\end{thebibliography}%

\end{document}